\documentclass[runningheads]{llncs}
\usepackage[T1]{fontenc}
\usepackage{graphicx,verbatim}
\usepackage{multirow} 
\usepackage{cite}
\usepackage{amsfonts, amsmath}
\usepackage[pagebackref,breaklinks,colorlinks]{hyperref}
\usepackage{color,xcolor,colortbl}
\usepackage[capitalize, noabbrev]{cleveref}
\usepackage{tabularx, booktabs}
\usepackage{xspace}
\begin{document}
\newcommand{\methodname}{RealDeal\xspace}
\title{\methodname: Enhancing \underline{Real}ism and \underline{De}t\underline{a}i\underline{l}s in Brain Image Generation via Image-to-Image Diffusion Models}

\author{
Shen Zhu\and
Yinzhu Jin\and
Tyler Spears \and
Ifrah Zawar\and
P. Thomas Fletcher
}

\authorrunning{S. Zhu et al.}
\institute{University of Virginia, Charlottesville, VA, USA \\
\email{\{sz9jt, yj3cz, tas6hh, emv5rf, ptf8v\}@virginia.edu}}
    
\maketitle              %
\begin{abstract}

Generative models have been widely adopted in the biomedical domain, especially in image generation applications. Latent diffusion models achieve state-of-the-art results in generating brain MRIs.
However, due to latent compression, generated images from these models are overly smooth, lacking fine anatomical structures and scan acquisition noise that are typically seen in real images.
We propose image-to-image diffusion models that are designed to enhance the realism and details of generated brain images by introducing sharp edges, fine textures, subtle anatomical features, and imaging noise.
This work formulates the realism enhancing and detail adding process as an image-to-image diffusion model, which refines the quality of LDM-generated images. 
We employ commonly used metrics like FID and LPIPS for image realism assessment.
Furthermore, we introduce new metrics to quantify the improved realism of images generated by \methodname in terms of image noise distribution, sharpness, and texture.

\keywords{Diffusion Model  \and Generative Model \and Brain}
\end{abstract}

\section{Introduction}
Deep generative models have demonstrated great capabilities in biomedical imaging~\cite{khader2023denoising, jin2024measuring, jinfeature, gu2023biomedjourney, zhu2025point, spears2025medil, shrivastava2023nasdm, bieder2024memory},
with applications for data augmentation~\cite{khader2023denoising, durrer2024denoising}, anomaly detection~\cite{bieder2024memory}, interpretability of machine learning models~\cite{jin2024measuring, jinfeature}, and counterfactual reasoning~\cite{gu2023biomedjourney, zhu2025point}. 
In particular, generating realistic brain MRIs~\cite{pinaya2022brain, khader2023denoising, spears2025medil, friedrich2024wdm} is relevant for both research and clinical use, where generated images can support deep learning models trained on limited datasets and improve diagnostic accuracy. 
However, generating high-fidelity images remains challenging due to the high dimensionality and complex anatomical structures present in brain MRIs.

Latent diffusion models (LDMs)~\cite{ho2020denoising, rombach2022high} address the problem by encoding 3D volumes into a lower-dimensional latent space. 
While the compression into the latent space reduces the computational overhead of the generative model, it also often diminishes or loses high-frequency content—such as sharp edges, fine textures, subtle anatomical features, and imaging noise—that are critical for accurate brain image representation.
As a result, images synthesized by LDMs tend to appear overly smooth and lack detailed features, as shown in Figure \ref{fig:qual_recon} and Figure \ref{fig:qual_synth}, potentially limiting their utility in clinical or research applications where precise anatomical fidelity is crucial.

Other strategies for high-resolution image generation have also been proposed.
Patch-based~\cite{ding2023patched, wang2023patch, jiang2025latent} generation divides a high-resolution image into tiles and generates the patches independently.
These methods are conditioned on global image encodings or text encodings to guide local generation. 
Bieder et al.~\cite{bieder2024memory} propose patch-based training of diffusion models to save computational overhead and use it to perform brain tumor segmentation from MRI.
However, to our knowledge, patch-based generation has not been extended to 3D or applied to biomedical imaging, where anatomical precision is critical.
To provide strong anatomical context for generating local details, we adopt LDM-generated volumes—which already capture global anatomical structure—as the conditioning input. Compared to global latent or text conditioning, the LDM output serves as a spatially-aligned and anatomically meaningful prior. Besides, with such a design, we can directly apply our method to existing LDM models for brain MRI generation as a post-hoc refinement module.

Thus, we propose \methodname, a two-step framework that enhances the realism and details of LDM-generated brain images.
\methodname integrates a patch-based image-to-image diffusion model~\cite{saharia2022palette} as a post-processing refinement step, explicitly conditioned on the initial LDM-generated images.
This additional step restores the high-frequency details lost in the compressed latent space, leading to improved image fidelity and anatomical accuracy. 
We further demonstrate the effectiveness of our approach compared to using LDM alone in improving image realism by using both qualitative comparison and quantitative analysis.
In our quantitative analysis, we employ commonly used metrics like FID~\cite{heusel2017gans}, and LPIPS~\cite{zhang2018perceptual} for image realism.
Furthermore, we also formulate novel metrics for measuring image noise distribution, sharpness, and texture of generated images.

\section{Enhancing Realism with RealDeal}

An LDM offers a two-stage framework for generating brain MRIs~\cite{khader2023denoising, pinaya2022brain}. First, images are compressed into a latent space using a convolutional autoencoder. Then, a diffusion model is trained to model and sample from the latent distribution. However, this compression often removes high-frequency content, resulting in overly smooth images that lack details for clinical and research applications.

\begin{figure}
    \centering
    \includegraphics[width=0.9\linewidth]{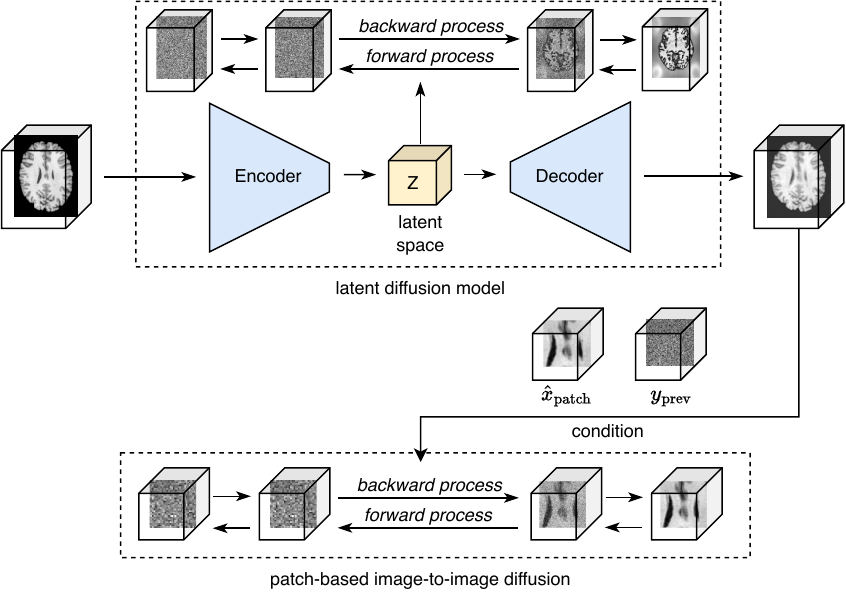}
    \caption{Model architecture. The output from LDM is used as condition to guide the patch-based image-to-image diffusion model.}
    \label{fig:architecture}
\end{figure}

To address this, we propose \methodname, which introduces an image-to-image diffusion model~\cite{saharia2022palette}
as a post-processing refinement step, conditioned on the initial coarse LDM output 
(Figure~\ref{fig:architecture}). This enhancement recovers sharp edges, fine textures, subtle anatomical features, and imaging noise
lost during compression. The following sections describe the LDM and refinement components in detail.

\subsection{Initial Generation Step: Latent Diffusion Model}

The two-stage training strategy of LDMs includes an autoencoder and a diffusion model in the latent space. 
The autoencoder maps inputs $x \in \mathbb{R}^{D\times H\times W}$ to latent codes $z = \mathcal{E}(x) \in \mathbb{R}^{d\times h\times w}$ via a convolutional encoder $\mathcal{E}$, with a decoder $\mathcal{D}$ reconstructing $\hat{x} = \mathcal{D}(z)$ under hybrid losses. LDMs~\cite{rombach2022high} then implement diffusion in this compressed space $\mathbb{R}^{d\times h\times w}$. 
This decoupled training strategy enables efficient synthesis of high-resolution 3D volumes by optimizing both space complexity and generative fidelity.

\subsection{Refinement Step: Image-to-Image Diffusion Model}

In \methodname, the refinement step is implemented as a patch-based image-to-image diffusion model~\cite{saharia2022palette} that restores the high-frequency details lost in LDM.
Image-to-image diffusion models~\cite{saharia2022palette} are conditional diffusion models~\cite{saharia2022image, chen2020wavegrad} with images as conditions.

\begin{figure}
    \centering
    \includegraphics[width=0.9\linewidth]{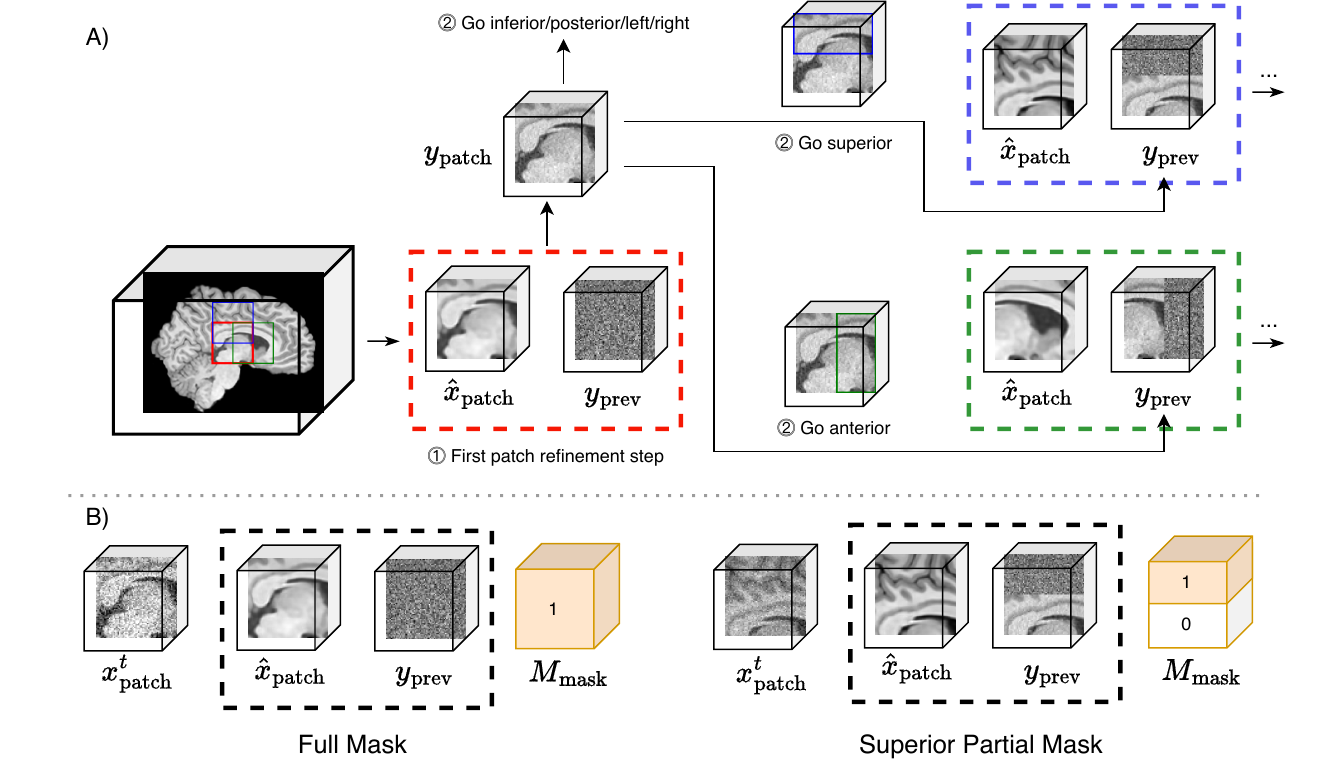}
    \caption{A) Patch-based refinement procedure for the whole image. B) Training data curation and mask types.}
    \label{fig:patch-based_and_mask-scheme}
\end{figure}

In our case, the model is of the form $p(y_\text{patch} \mid \hat{x}_\text{patch}, y_\text{prev})$, where $\hat{x}_\text{patch}$ and $y_\text{prev}$ are the conditions.
The refinement procedure always starts with the center patch, and iteratively progresses to the neighboring patches until we traverse all patches in an image.
During this iterative process, $\hat{x}_\text{patch}$ is a patch of the coarse image output from the LDM~\cite{rombach2022high}, and $y_\text{prev}$ is a partially restored patch from the previous refinement step, as illustrated in Figure \ref{fig:patch-based_and_mask-scheme}. 
One special case for this procedure is the first step, the center patch, where we do not have any previously restored patches. We set $y_\text{prev}$ to pure noise for the first step.

The condition $\hat{x}_\text{patch}$ provides the model with coarse anatomical shape and spatial context, while the condition $y_\text{prev}$ supplies the refined output from the previous patch. Together, these conditions ensure spatial coherence across patches and help prevent grid-like artifacts in the generated image. At the end of the refinement procedure, we get a whole refined brain MRI denoted as $y_\text{refined}$.

The model takes random Gaussian noise as input and is conditioned on $\hat{x}_\text{patch}$ and $y_\text{prev}$. It then progressively refines the input through an iterative denoising procedure.
At each diffusion step, the network leverages the structural guidance from the coarse image patch $\hat{x}_\text{patch}$ and information from the last refinement step $y_\text{prev}$ by explicitly concatenating them to the input and conditioning on them, thereby learning to predict and restore the residual high-frequency components, such as sharp edges, fine textures, subtle anatomical boundaries, and imaging noise.

During training, an original brain image patch $x_\text{patch}$ is first perturbed into a noisy image patch $x_\text{patch}^t$ with noise $\xi$, using a predefined Gaussian noise schedule. Its corresponding coarse image patch $\hat{x}_\text{patch}$, produced by the first stage, is provided as conditioning input. In addition, $y_\text{prev}$ is also provided as guidance and is computed as: $y_\text{prev} = x_\text{patch} * (1 - M_\text{mask}) + G_\text{noise} * M_\text{mask}$,
where $G_\text{noise}$ is random Gaussian noise with the same shape as $x_\text{patch}$, and $M_\text{mask}$ is a binary mask that can be full or partial (anterior, posterior, inferior, superior, left, and right) as shown in Figure \ref{fig:patch-based_and_mask-scheme}. During training, we use full masks $10\%$ of the time and partial masks $90\%$ of the time.

We train a neural network $f_{\psi}$, parameterized by $\psi$, to reverse this forward noise process by predicting the noise residual $\xi$, conditioned on $\hat{x}_\text{patch}$ and $y_\text{prev}$. This approach ensures that the model not only preserves the global anatomical structure captured by the LDM but also learns to infuse the refined images with the missing intricate details using guidance from $y_\text{prev}$. The training objective aims to minimize the discrepancy between the noise $\xi$ and the noise predicted by the network $f_{\psi}$, and we only calculate loss where $M_\text{mask} = 1$:

\begin{equation}
    \mathcal{L}_{\text{diff}} = \mathbb{E}_{t, x_\text{patch},\xi, \hat{x}_\text{patch}, y_\text{prev}}  
    \left\| [\xi - f_\psi(x_\text{patch}^t,t, \hat{x}_\text{patch}, y_\text{prev})] \times M_\text{mask} \right\|^2.
\end{equation}

\section{Experiments}
We designed a series of experiments to assess the effectiveness of \methodname in enhancing the realism and details of LDM-generated brain images. 
First, we qualitatively analyze how \methodname improves image quality for both reconstructed images and fully synthetic images. Besides, to complement the qualitative assessment, we employ commonly used metrics like FID~\cite{heusel2017gans}, and LPIPS~\cite{zhang2018perceptual} for image realism.
We further develop various quantitative metrics for different image quality aspects, like \textit{noise distribution}, \textit{sharpness}, and \textit{texture}.

The following sections describe the dataset, implementation details, evaluation metrics, and results.

\subsection{Dataset and Implementation Details}

We conduct our experiments using the Human Connectome Project (HCP) dataset~\cite{VANESSEN201362}. The HCP provides high-resolution, high-quality brain MRI scans, and we utilize T1-weighted structural MRI scans specifically. 

We use 1,113 subjects, splitting them into 80\% for training (890 subjects) and 20\% for testing (223 subjects). All images are cropped to a size of $[224, 288, 224]$ to ensure full brain coverage, with an isotropic voxel size of $0.7 \text{ mm}^3$. Image intensities are rescaled to the range $[-1, 1]$ using predefined global thresholds.

For LDM, we downsample the input $4\times$ in each dimension, with latent space size $[4, 56, 72, 56]$. For patch-based refinement, we use a patch size of $64^3$. For both latent diffusion and image-to-image diffusion, we use $1000$ timesteps with linear noise schedule, with batch sizes of $4$ and $32$ respectively. During the image-to-image model training, we use  $10\%$ full masks and $90\%$ partial masks.

\subsection{Quantitative Metrics and Results}

For \textit{reconstructed images}, we employ the widely adopted LPIPS~\cite{zhang2018perceptual} for perceptual similarity.
between the original images and their corresponding reconstructed and refined outputs.
We also systematically evaluate image quality with respect to three criteria: \textit{noise distribution}, \textit{sharpness}, and \textit{texture}. 

\textbf{LPIPS.} For each original, reconstructed, and refined images triplet, we sample 1,000 patch locations with a patch size of $64^2$ in each dimension. We then compute LPIPS~\cite{zhang2018perceptual} with AlexNet~\cite{krizhevsky2012imagenet}.
We use 20 original test images in total, and the results are summarized in Table \ref{tab:lpips}. 
It shows that the refined images consistently exhibit lower feature distances to the original images than the reconstructed ones, indicating improved local structural fidelity.

\begin{table*}[t]
    \centering\small
    \setlength{\tabcolsep}{2pt}
    \caption{LPIPS, textual similarity, and KL divergence between original images and their corresponding reconstructed and \methodname-refined images.}
    \begin{tabular}{l ccc ccc ccc l}
    
        \toprule
        & \multicolumn{3}{c}{LPIPS $\downarrow$}
        & \multicolumn{3}{c}{\begin{tabular}{@{}c@{}}Whole Brain \\ Textual Similarity $\downarrow$\end{tabular}}
        & \multicolumn{3}{c}{\begin{tabular}{@{}c@{}}Cerebellum \\ Textual Similarity $\downarrow$\end{tabular}}
        & \multirow{3}{*}{KL $\downarrow$}
        \\
        
	\cmidrule(lr){2-4}
        \cmidrule(lr){5-7}
        \cmidrule(lr){8-10}
  
	& \multicolumn{1}{c}{dim0} & dim1 & dim2
        & dim0 & dim1 & dim2
        & dim0 & dim1 & dim2\\
  
	\midrule
  
	Recon.
        & 0.078 & 0.073 & 0.079
        & 7.602 & 6.486 & 7.181
        & 2.639 & 3.243 & 3.149
        & 0.775
        \\
            
        \methodname  
        & \textbf{0.029} & \textbf{0.027} & \textbf{0.030}
        & \textbf{6.451} & \textbf{5.610} & \textbf{6.093}
        & \textbf{2.143} & \textbf{2.734} & \textbf{2.715}
        & \textbf{0.149}
        \\
        
	\bottomrule
	\end{tabular}
	\label{tab:lpips}
\end{table*}

\textbf{Noise Distribution.} We first extract noise from images using the ANTS software suite~\cite{manjon2010adaptive}, and then we compare the noise distributions in the white matter region of the original images to those of their corresponding reconstructed and refined images using KL divergence. Lower KL divergence to the original noise distribution indicates better preservation of the imaging noise properties. We summarized the results in Table \ref{tab:lpips}. Images refined by \methodname have much more similar distribution of noise to the original images than LDM-reconstructed images.

\textbf{Image Sharpness.} For sharpness, we propose to use patch-based \textit{Laplacian variance}, which measures the local intensity variations by computing the variance of the Laplacian response within each patch. Higher variance values correspond to sharper image regions with more edge information. 
For each original, reconstructed, and refined images triplet, we smooth the images with Gaussian filter, with $\sigma=0.5$ voxels, in order to be robust to noise.
We then sample 1,000 patch locations with patch size $64^3$ and compute their Laplacian variance. The results are shown in Table \ref{tab:sharpness}. The original images show the most sharpness, closely followed by refined images, which is in conformity with qualitative analysis.

\begin{table*}[t]
    \centering\small
    \setlength{\tabcolsep}{2pt}
    \caption{Sharpness results for original, reconstructed, and \methodname-refined images.}
    \begin{tabular}{l ccc}
        \toprule
		
        & Orig. & Recon. & \methodname\\
  
	\midrule
  
	{Sharpness $\uparrow$}
        & 0.0139 & 0.0063 & 0.0111\\

	\bottomrule
	\end{tabular}
	\label{tab:sharpness}
\end{table*}

\textbf{Textual Similarity.} 
While Laplacian variance captures overall image sharpness, we complement this with a Histogram of Oriented Gradients (HOG)-based metric to assess local textural similarity and structural detail preservation.
We develop this metric by computing HOG descriptors in the whole brain and cerebellum. 
The textural similarity is then computed as the $L^2$ distance between HOG descriptors, quantifying local gradient structure similarity. 
We summarize the HOG results in Table \ref{tab:lpips}. Our refined images show higher textual similarity than the reconstructed ones in both the whole brain and the cerebellum.

For \textit{fully synthetic images}, we adopt commonly used metrics like FID \cite{heusel2017gans}, coverage, and density~\cite{naeem2020reliable} to measure how close the distributions of LDM-generated images and \methodname-refined images to that of the original images. Coverage and density metrics~\cite{naeem2020reliable} are computed based on how frequently the generated samples fall within the $k$-nearest neighbor spheres of real data. Specifically, coverage is measured as the percentage of real data points whose neighborhoods are hit, while density reflects the average number of hits per neighborhood. We set $k=10$, we then sample 20 LDM-generated images, and their corresponding \methodname-refined images. Subsequently, we follow the same patch-based strategy used by LPIPS. We show the results in Table \ref{tab:synthetic}.

\begin{table*}[t]
    \centering\small
    \setlength{\tabcolsep}{2pt}
    \caption{FID, coverage, and density results for fully synthetic images by LDM and their corresponding \methodname-refined images compared to original images. Bottom row shows the results between original images.}
    \begin{tabular}{l ccc ccc ccc}
        \toprule

        & \multicolumn{3}{c}{FID $\downarrow$}
        & \multicolumn{3}{c}{\begin{tabular}{@{}c@{}}Coverage $\uparrow$ \\ $k=10$\end{tabular}}
        & \multicolumn{3}{c}{\begin{tabular}{@{}c@{}}Density $\uparrow$ \\ $k=10$\end{tabular}}\\
        
	\cmidrule(lr){2-4}
        \cmidrule(lr){5-7}
        \cmidrule(lr){8-10}
  
	& \multicolumn{1}{c}{dim0} & dim1 & dim2
        & dim0 & dim1 & dim2
        & dim0 & dim1 & dim2\\
  
	\midrule
  
	Synthetic
        & 46.62 & 48.77 & 46.59
        & 0.311 & 0.322 & 0.310
        & 0.172 & 0.189 & 0.168\\
            
        \methodname  
        & \textbf{17.30} & \textbf{18.23} & \textbf{19.17}
        & \textbf{0.763} & \textbf{0.738} & \textbf{0.745}
        & \textbf{0.492} & \textbf{0.462} & \textbf{0.494}\\

        \midrule

        \textcolor{gray}{Orig.}
        & \textcolor{gray}{2.48} & \textcolor{gray}{2.74} & \textcolor{gray}{2.84}
        & \textcolor{gray}{0.971} & \textcolor{gray}{0.970} & \textcolor{gray}{0.936}
        & \textcolor{gray}{0.910} & \textcolor{gray}{0.912} & \textcolor{gray}{0.907}\\
        
	\bottomrule
	\end{tabular}
	\label{tab:synthetic}
\end{table*}

\subsection{Qualitative Results}

\begin{figure}
    \centering
    \includegraphics[width=1.0\linewidth]{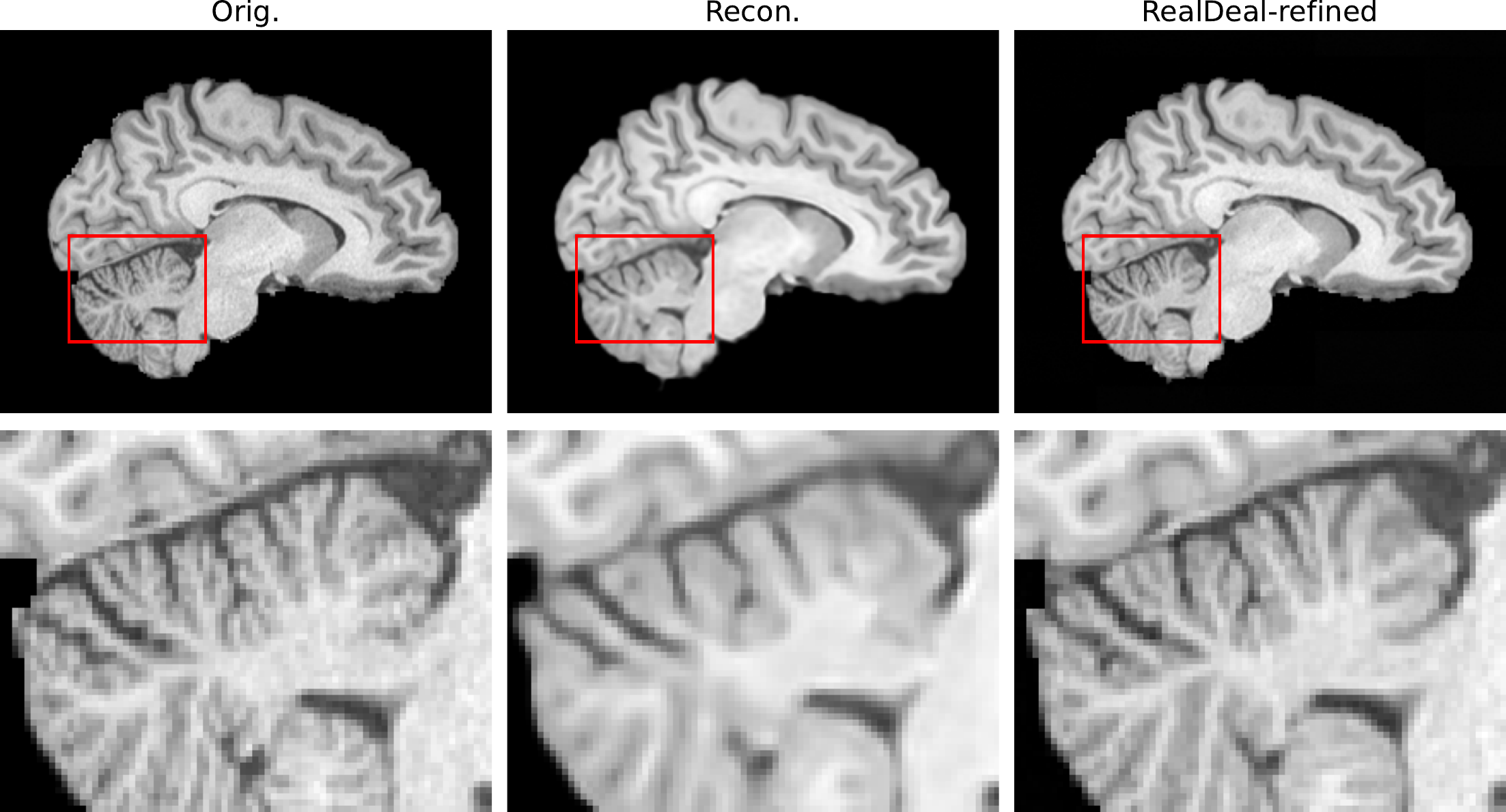}
    \caption{Qualitative results for an original sagittal slice and its corresponding autoencoder-reconstructed and \methodname-refined slices.}
    \label{fig:qual_recon}
\end{figure}

We first qualitatively analyze how the LDM model processes real brain MRI scans by examining the autoencoder-reconstructed images, and their corresponding \methodname-refined images to determine the extent of detail restoration.
We show a sagittal slice of the original, reconstructed, and refined images of one subject in Figure \ref{fig:qual_recon}. 
We observe that the refined image has sharper edge content than the reconstructed image. Besides, the refined image exhibits more realistic textual and imaging noise, where the reconstructed image is overly smooth. 

\begin{figure}
    \centering
    \includegraphics[width=0.7\linewidth]{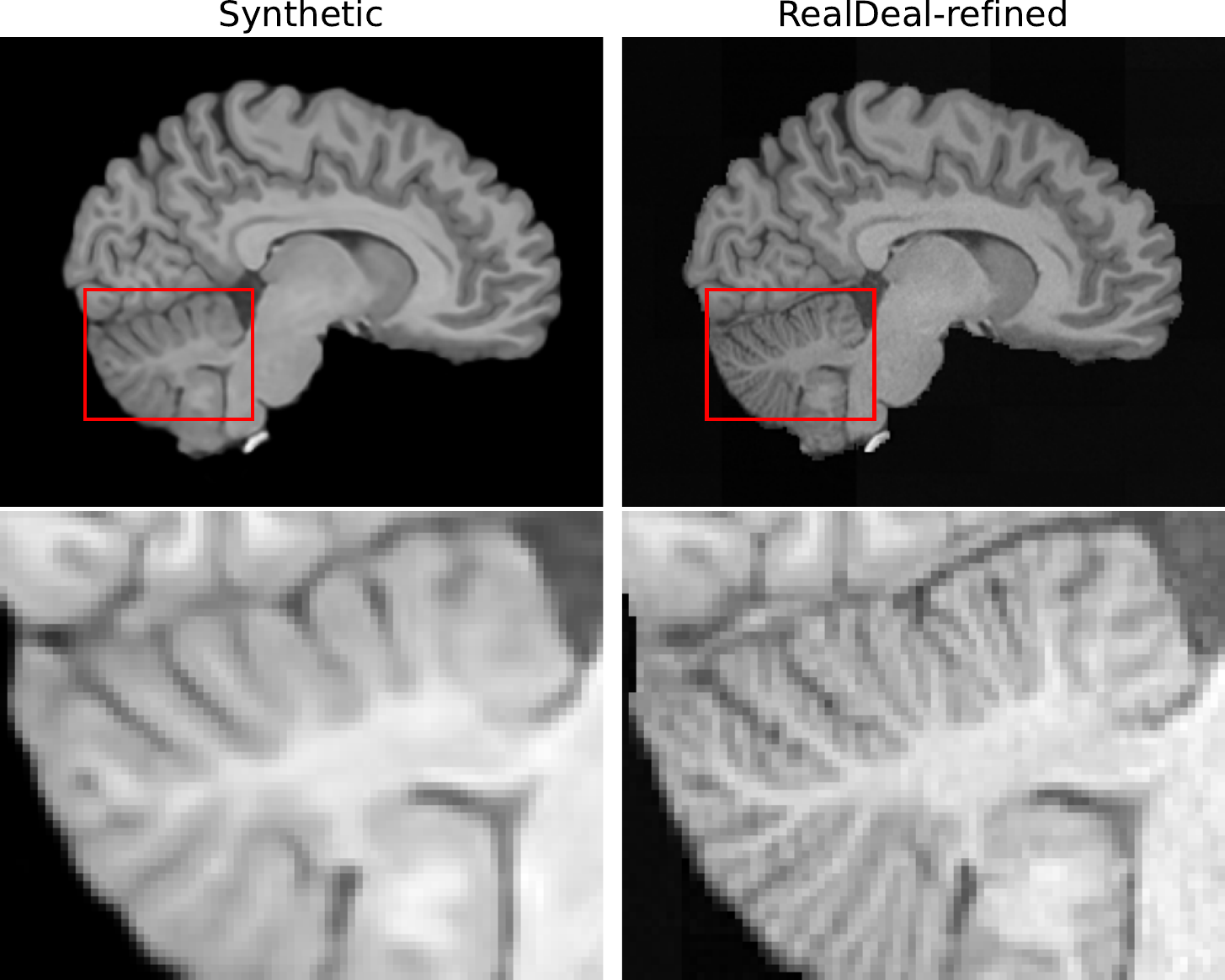}
    \caption{Qualitative results for an LDM-generated sagittal slice and its \methodname-refined slice.}
    \label{fig:qual_synth}
\end{figure}

Besides, we also evaluate the full generative process by applying \methodname to images synthesized by the LDM model, assessing its ability to restore fine anatomical structures and improve overall image fidelity for synthetic images. We plot a representative sagittal slice of one synthesized image, and a slice of the corresponding refined image in Figure \ref{fig:qual_synth}. The refined image also shows more resemblance to real data than the synthesized image in terms of image noise, sharpness, and texture.

These qualitative experiments allow us to visualize the loss of high-frequency details introduced by the latent space compression, and examine the effectiveness of \methodname in enhancing image realism and restoring realistic image details.

\section{Conclusion}

In this work, we presented \methodname, an image-to-image diffusion framework that enhances the realism of brain images generated by LDMs~\cite{rombach2022high}. By addressing the loss of high-frequency details inherent to latent space compression, \methodname refines coarse LDM outputs to recover fine anatomical structures, realistic textures, and imaging noise. We validated its effectiveness through qualitative results and quantitative metrics, including FID~\cite{heusel2017gans}, LPIPS~\cite{zhang2018perceptual}, and newly proposed measures targeting sharpness, noise, and texture.
Compared to LDM-generated images, \methodname produces outputs with improved anatomical fidelity and visual realism. As a refinement module, it can be integrated into existing generative pipelines, with future extensions planned for multi-modal and clinical imaging.
By enhancing anatomical fidelity and realistic imaging features, RealDeal may enable more robust diagnostic and prognostic tools in data-scarce clinical settings and has the potential to transform clinical care in neurological disorders.

\section{Acknowledgments}
This research received funding support from Alzheimer's Association (Grant Number: AACSFD-22-974008, PI: Dr. Zawar). This work was partially supported by NSF Smart and Connected Health grant 2205417.

\newpage

\bibliographystyle{splncs04}
\bibliography{ref}

\newpage
\setcounter{section}{0}  %

\begin{center}
    {\LARGE \bfseries \methodname: Supplement Material \par}
\end{center}

\section{Training Details}

We use 1,113 subjects, splitting them into 80\% for training (890 subjects) and 20\% for testing (223 subjects). All images are cropped to a size of $[224, 288, 224]$. Image intensities are rescaled to the range $[-1, 1]$ using predefined global thresholds.

For the autoencoder of the LDM, we train a 3D ResNet-based autoencoder using the AdamW optimizer with a learning rate of 1e-3 and weight decay of 1e-5. The encoder consists of convolutional and ResNet blocks with two downsampling stages, reducing spatial resolution by a factor of 4. The bottleneck maps features to a low-dimensional representation using $1\times1\times1$ convolutions. The decoder mirrors the encoder with upsampling layers and residual blocks, reconstructing the input volume. LeakyReLU is used as the activation function, and the output is scaled to $[-1, 1]$ using a final sigmoid layer.

We use a 3D latent diffusion model trained in the latent space of the LDM. The U-Net backbone has 3 spatial levels with increasing channels $[256, 512, 768]$, using two residual blocks per level and self-attention at the last two scales. The model operates on 4-channel latent inputs and outputs, with no external conditioning. We use a linear beta noise schedule with 1,000 steps, $\beta_\text{start} = 0.0015$, $\beta_\text{end} = 0.0205$, and perform velocity prediction. The model is trained for 750 epochs using a learning rate of 2.5e-5 and batch size of 4 on 3D latent volumes of shape $56 \times 72 \times 56$.

We trained our 3D patch-based image-to-image diffusion model for 2,700 epochs using a batch size of 32. The model architecture consists of a 3D U-Net with three resolution levels, using 128, 256, and 256 channels respectively, and self-attention applied at the final level. Each level includes two residual blocks. The model was trained with a base learning rate of $1 \times 10^{-5}$ and a noise scheduler using a scaled linear beta schedule with 1,000 diffusion steps ($\beta_\text{start}=0.0015$, $\beta_\text{end}=0.0205$). We use $10\%$ full masks and $90\%$ partial masks during training.

\section{More Quantitative Results}

For noise distribution, we provide additional KL divergence results in the lateral ventricle region. Lower KL divergence to the original noise distribution indicates better preservation of the imaging noise properties. We summarized the results in Table \ref{tab:ventricle}.

\begin{table*}[t]
    \centering\small
    \setlength{\tabcolsep}{2pt}
    \caption{KL divergence between original images and their corresponding reconstructed and \methodname-refined images in the lateral ventricle regions.}
    \begin{tabular}{l l}
    
        \toprule
        & {KL $\downarrow$}
        \\
  
	\midrule
  
	Recon.
        & 0.3941
        \\
            
        \methodname  
        & \textbf{0.0702}
        \\
        
	\bottomrule
	\end{tabular}
	\label{tab:ventricle}
\end{table*}

For \textit{fully synthetic images}, calculate coverage, and density to measure how close the distributions of LDM-generated images and \methodname-refined images to that of the original images. Coverage and density metrics are computed based on how frequently the generated samples fall within the $k$-nearest neighbor spheres of real data. We use other $k$ values where $k=5, 20$ and provide the results in Table \ref{tab:k5res} and Table \ref{tab:k20res}.

\begin{table*}[t]
    \centering\small
    \setlength{\tabcolsep}{2pt}
    \caption{Coverage, and density results for fully synthetic images by LDM and their corresponding \methodname-refined images compared to original images. $k=5$}
    \begin{tabular}{l ccc ccc}
        \toprule
        
        & \multicolumn{3}{c}{\begin{tabular}{@{}c@{}}Coverage $\uparrow$ \\ $k=5$\end{tabular}}
        & \multicolumn{3}{c}{\begin{tabular}{@{}c@{}}Density $\uparrow$ \\ $k=5$\end{tabular}}\\
        
	\cmidrule(lr){2-4}
        \cmidrule(lr){5-7}
  
	& \multicolumn{1}{c}{dim0} & dim1 & dim2
        & dim0 & dim1 & dim2\\
  
	\midrule
  
	Synthetic
        & 0.192 & 0.196 & 0.186
        & 0.150 & 0.165 & 0.143\\
            
        \methodname  
        & \textbf{0.588} & \textbf{0.558} & \textbf{0.572}
        & \textbf{0.453} & \textbf{0.422} & \textbf{0.452}\\

        \midrule

        \textcolor{gray}{Orig.}
        & \textcolor{gray}{0.916} & \textcolor{gray}{0.910} & \textcolor{gray}{0.884}
        & \textcolor{gray}{0.880} & \textcolor{gray}{0.878} & \textcolor{gray}{0.867}\\
        
	\bottomrule
	\end{tabular}
	\label{tab:k5res}
\end{table*}

\begin{table*}[t]
    \centering\small
    \setlength{\tabcolsep}{2pt}
    \caption{Coverage, and density results for fully synthetic images by LDM and their corresponding \methodname-refined images compared to original images. $k=20$}
    \begin{tabular}{l ccc ccc}
        \toprule
        
        & \multicolumn{3}{c}{\begin{tabular}{@{}c@{}}Coverage $\uparrow$ \\ $k=20$\end{tabular}}
        & \multicolumn{3}{c}{\begin{tabular}{@{}c@{}}Density $\uparrow$ \\ $k=20$\end{tabular}}\\
        
	\cmidrule(lr){2-4}
        \cmidrule(lr){5-7}
  
	& \multicolumn{1}{c}{dim0} & dim1 & dim2
        & dim0 & dim1 & dim2\\
  
	\midrule
  
	Synthetic
        & 0.468 & 0.482 & 0.471
        & 0.203 & 0.216 & 0.202\\
            
        \methodname  
        & \textbf{0.882} & \textbf{0.870} & \textbf{0.855}
        & \textbf{0.531} & \textbf{0.499} & \textbf{0.529}\\

        \midrule

        \textcolor{gray}{Orig.}
        & \textcolor{gray}{0.975} & \textcolor{gray}{0.975} & \textcolor{gray}{0.939}
        & \textcolor{gray}{0.933} & \textcolor{gray}{0.933} & \textcolor{gray}{0.927}\\
        
	\bottomrule
	\end{tabular}
	\label{tab:k20res}
\end{table*}

\clearpage

\section{More Qualitative Results}

\subsection{Original, autoencoder-econstructed, and \methodname-refined images of more subjects}

\begin{figure}
    \centering
    \includegraphics[width=0.9
    \linewidth]{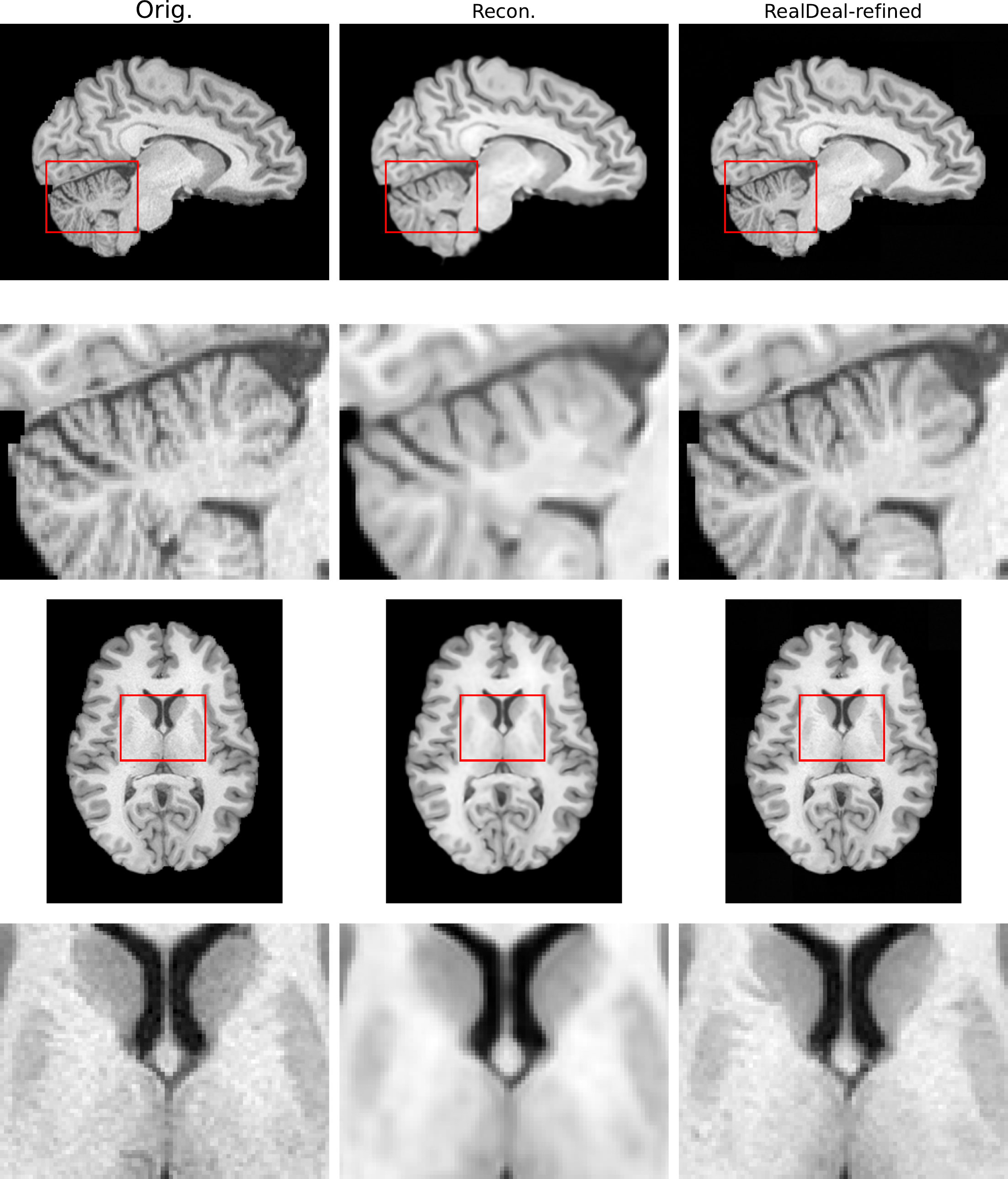}
    \caption{Subject 0 original image and its corresponding autoencoder-reconstructed and \methodname-refined images.}
    \label{fig:sample0}
\end{figure}

\begin{figure}
    \centering
    \includegraphics[width=0.9\linewidth]{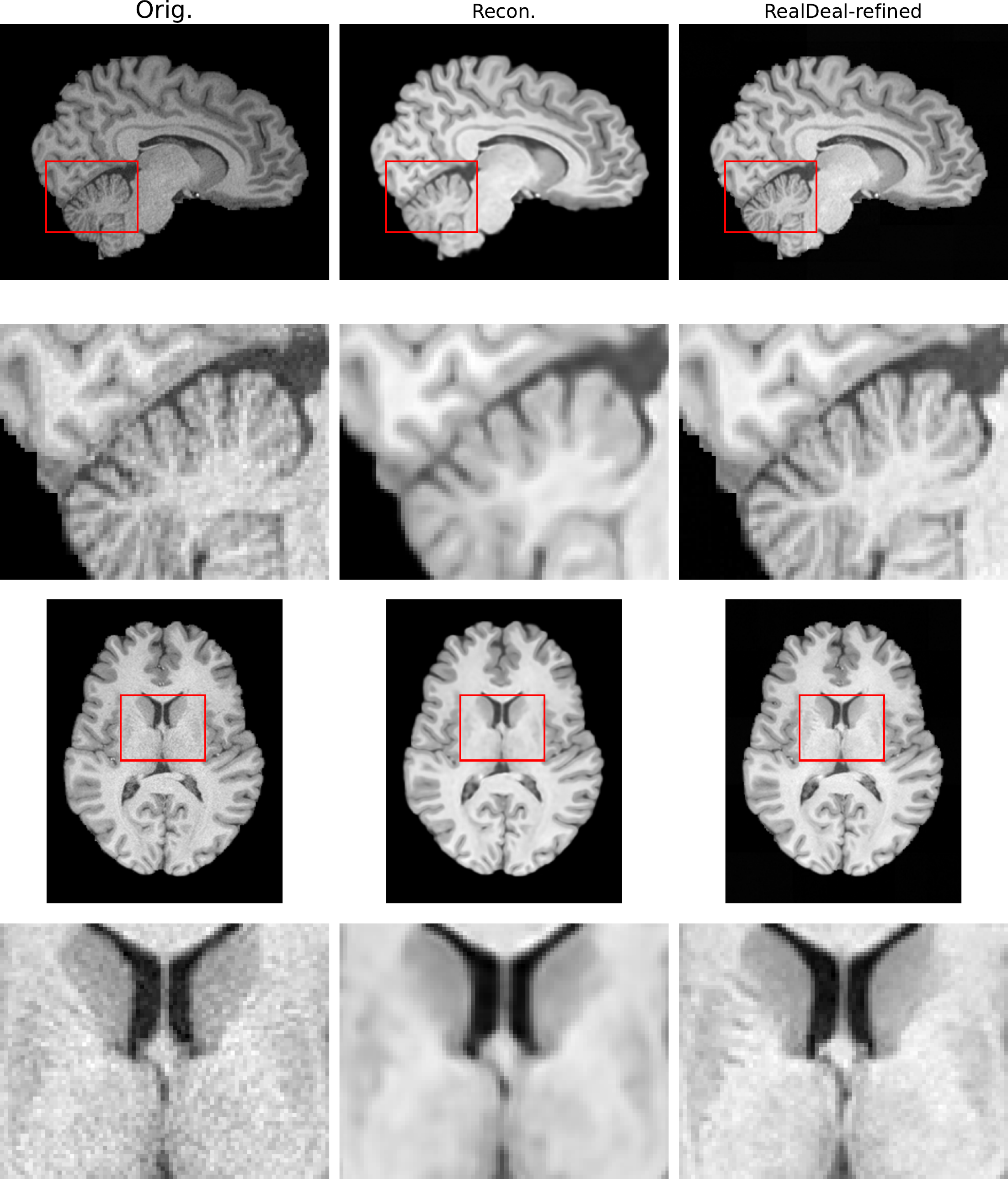}
    \caption{Subject 1 original image and its corresponding autoencoder-reconstructed and \methodname-refined images.}
    \label{fig:sample4}
\end{figure}

\begin{figure}
    \centering
    \includegraphics[width=0.9\linewidth]{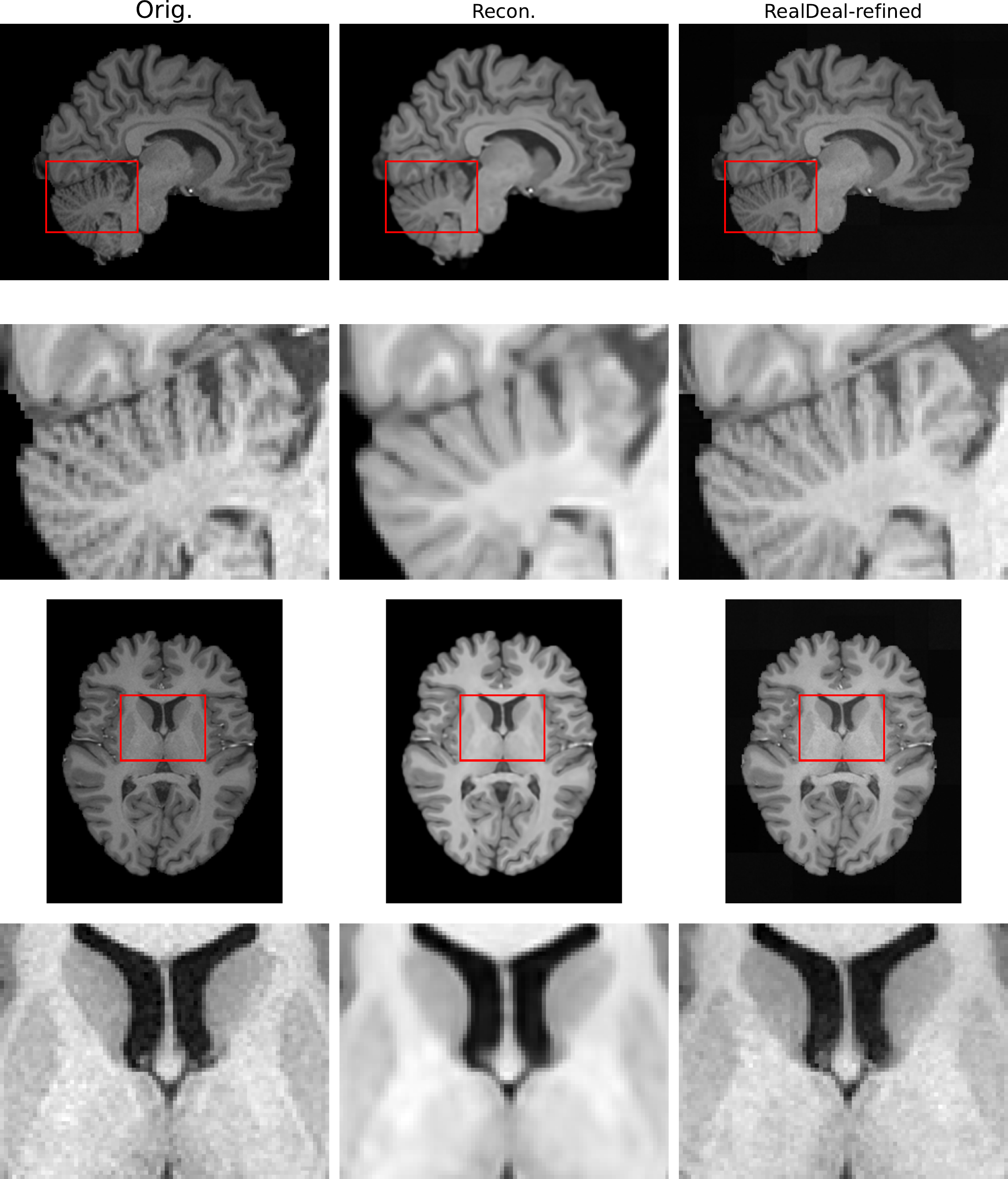}
    \caption{Subject 2 original image and its corresponding autoencoder-reconstructed and \methodname-refined images.}
    \label{fig:sample15}
\end{figure}

\clearpage

\subsection{More LDM-generated samples and their corresponding \methodname-refined images}

\begin{figure}
    \centering
    \includegraphics[width=1\linewidth]{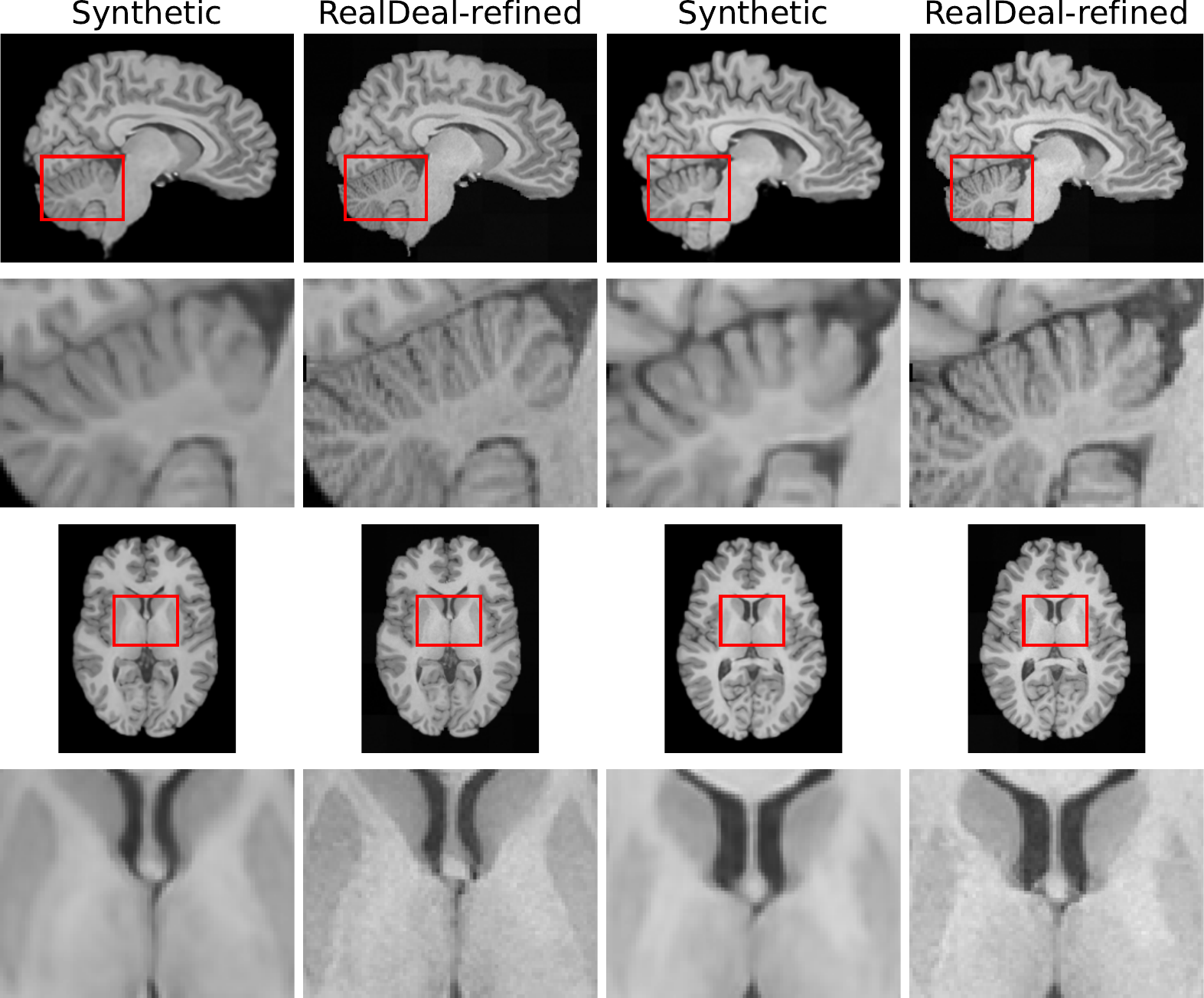}
    \caption{Qualitative results of two LDM-generated samples and their \methodname-refined images.}
    \label{fig:synth89}
\end{figure}

\begin{figure}
    \centering
    \includegraphics[width=1\linewidth]{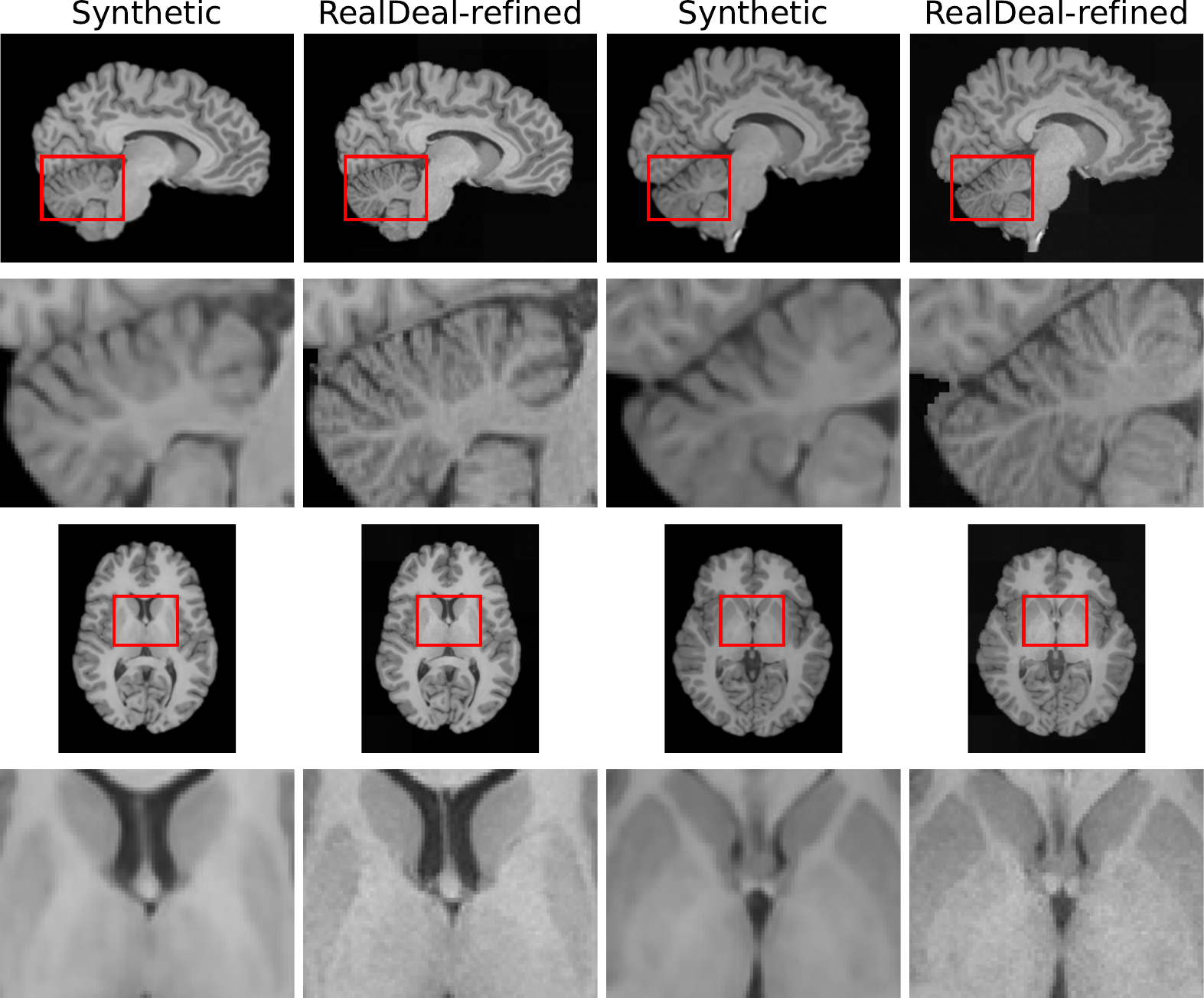}
    \caption{Qualitative results of two LDM-generated samples and their \methodname-refined images.}
    \label{fig:synth23}
\end{figure}

\begin{figure}
    \centering
    \includegraphics[width=1\linewidth]{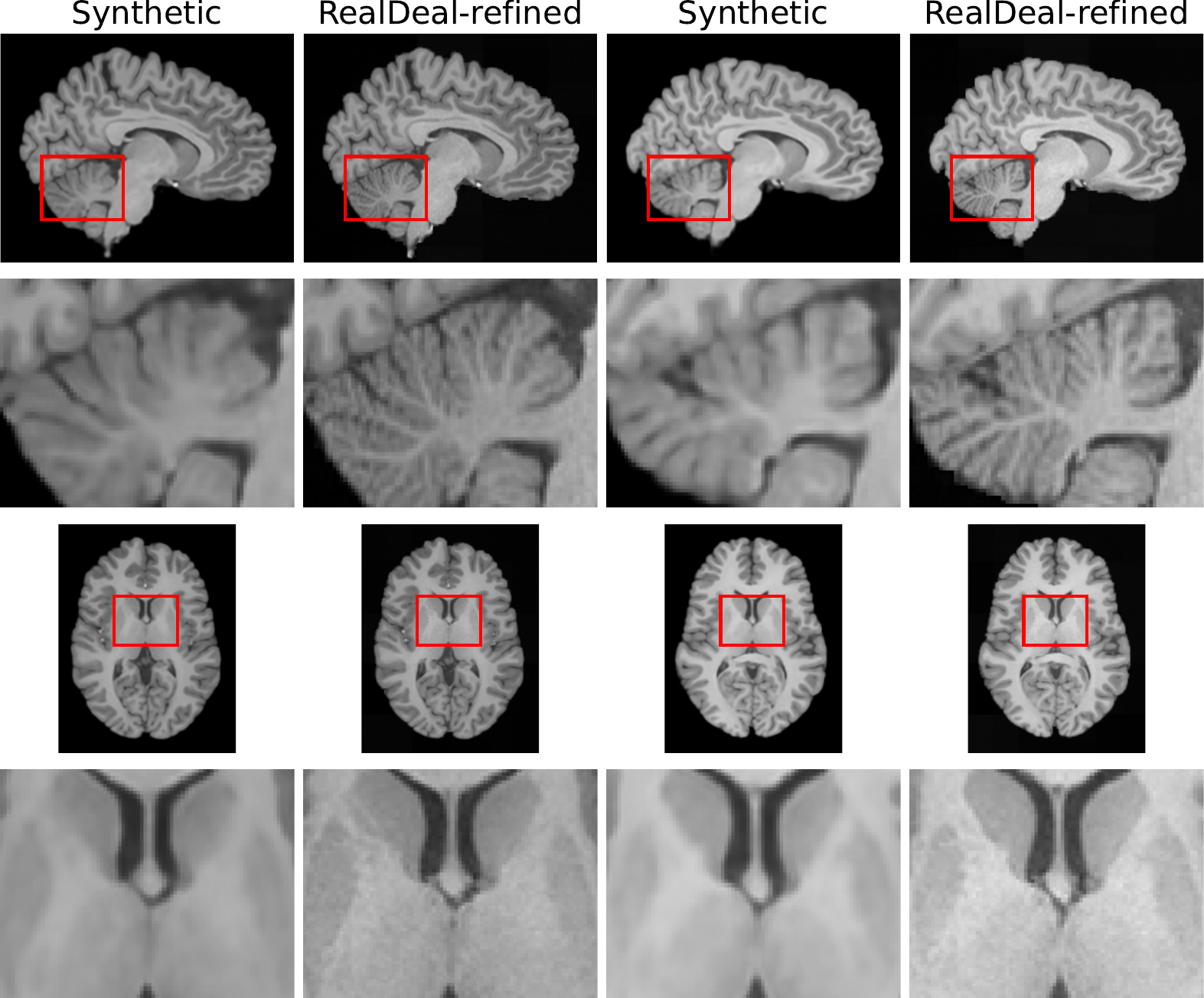}
    \caption{Qualitative results of two LDM-generated samples and their \methodname-refined images.}
    \label{fig:synth1617}
\end{figure}

\end{document}